\documentclass[%
 reprint,
showpacs,
 amsmath,amssymb,
 aps,
]{revtex4-1}

\usepackage{graphicx}
\usepackage{dcolumn}
\usepackage{bm}
\usepackage{physics}
\usepackage{float}
\usepackage[]{hyperref}

\newcommand{\arxiv}[1]{arXiv:\href{http://www.arxiv.org/abs/#1}{#1}}

\def\dm{\Delta m^2}

\def\dmt{\Delta \tilde m^2}

\newcommand{\asym}[1]{\mathcal{A}^\mathrm{#1}}

\begin{document}


\title{
		Disentangling genuine from matter-induced CP violation
		in neutrino oscillations
}

\author{Jos\'e Bernab\'eu}
\author{Alejandro Segarra}%
\affiliation{%
	Department of Theoretical Physics, University of Valencia
	and IFIC, Univ.~Valencia - CSIC, Burjassot, Valencia, Spain
}%

\date{\today}

\begin{abstract}
	We prove that, in any flavor transition, neutrino oscillation CP violating
	asymmetries in matter have two disentangled components:
	i) a CPT-odd T-invariant term, non-vanishing iff there are interactions
	with matter; 
	ii) a T-odd CPT-invariant term, non-vanishing iff there is genuine CP violation. 
	As function of the baseline,
	these two terms are distinct $L$-even and $L$-odd observables
	to separately test 
	(i) matter 	effects sensitive to the neutrino hierarchy and 
	(ii) genuine CP violation in the neutrino sector. 
	For the golden $\nu_\mu \to \nu_e$ channel,
	the different energy distributions of the two components
	provide a signature of their separation.
	At long baselines, they show oscillations
	in the low and medium energy regions,
	with zeros at different positions and peculiar behavior around the zeros.
	We discover a
	magic energy $E = (0.91 \pm 0.01)$ GeV at $L = 1300$ km with vanishing 
	CPT-odd component
	and maximal genuine CP asymmetry proportional to
	$\sin\delta$, with $\delta$ the weak CP phase.
	For energies above $1.5$ GeV, the sign of the CP asymmetry discriminates
	the neutrino hierarchy.
\end{abstract}

\maketitle


The last two decades have seen a revolution in neutrino physics
with the discovery of, and precision studies on, flavor oscillations
in atmospheric~\cite{atmospheric}, solar~\cite{solar}, 
reactor~\cite{reactor} and accelerator~\cite{accelerator} neutrinos. 
These phenomena are interpreted in terms of non-vanishing masses and flavor mixing, 
the unitary PMNS matrix describing the mismatch between flavor and mass eigenstates. 
Global fits to all observable quantities provide better and better determination 
of the two mass differences $\dm_{21}$ and $\abs{\dm_{31}}$ 
and the three mixing angles~\cite{mariam, nufit, fogli}. 
Besides the pending fundamental questions on the Dirac-Majorana nature of neutrinos 
and their absolute mass scale, studied by means of other methods, 
neutrino flavor oscillations have novel challenges 
for the next-generation experiments like T2HK~\cite{HK} and DUNE~\cite{DUNE}.
Above all, once known that the three mixing angles are non-vanishing~\cite{T2K, DAYABAY, RENO, 2CHOOZ}, 
they should answer whether the lepton sector of elementary particles 
also incorporates CP violation, 
opening the door to concepts able to explain the matter-antimatter asymmetry of the Universe
through leptogenesis~\cite{baryogenesis} at higher energy scales. 
A second open problem is the ordering of the neutrino mass spectrum, 
with  the so-called Normal or Inverted Hierarchies associated 
to the positive or negative 
sign of $\dm_{31}$.

Long baseline neutrino oscillation experiments propagate neutrinos from the source, 
created as muonic flavor, 
to the detector through the Earth mantle. 
The observation of CP violation needs an appearance experiment to a different flavor,
and the ``suppressed'' transition to the electronic flavor is favored. 
The corresponding CP violation asymmetry, 
defined in terms of the transition probabilities for neutrinos and antineutrinos 
$
	\asym{CP}_{\alpha\beta} \equiv
	P (\nu_\alpha \to \nu_\beta) -
	P (\bar \nu_\alpha \to \bar \nu_\beta)\,,
$
is an odd function of $L/E$, with $L$ the baseline and $E$ the relativistic neutrino energy,
iff the propagation takes place in vacuum.
Independent of particular theoretical frameworks, 
this observable is a bona-fide direct proof of CP violation. 
However, in actual experiments the propagation takes place in matter, which is CP-asymmetric,
and a fake CP violation is originated through the different interaction
of electron neutrinos and antineutrinos with the electron density of ordinary
matter~\cite{MSW-W, MSW-MS}.
This complication in the quest for a direct evidence of fundamental CP violation is
widely recognized, and some observables~\cite{obs:donini,obs:akhmedov, obs:ohlsson}
have been tried for its separation.
On the other hand, 
matter-induced terms are
welcome in order to obtain information on the neutrino mass hierarchy. 
Due to this combined effect, the generalized attitude in the
scientific community has been 
to extract the CP phase $\delta$ in the $U_\mathrm{PMNS}$ mixing matrix from the global fits:
a resulting value different from 0 or $\pi$ is taken as evidence of CP violation. 
Such a methodology is, however, not guaranteeing that 
the experiment did actually observe CP violation
---any quantity 
sensitive to $\delta$ would make this job,
as happens in bare transition probabilities due to 
the CP-conserving $\cos\delta$ terms.

The present paper represents 
the restoration of the idea
that a direct evidence of symmetry violation means 
the measurement in a single experiment 
of an observable odd under the symmetry. 
The Concept exploited here is based on the fact that 
genuine and matter-induced CP violation have opposite behaviors~\cite{Banuls}
under the other discrete symmetries of Time-Reversal T and CPT: 
whereas genuine CP violation is odd under T and even under CPT, 
the matter effect is T-even and CPT-odd. 
Although they are well separated in the effective Hamiltonian,
in general they are not in the experimental observables and, in particular, in the CP asymmetry. 
The ideal way to solve this problem would be 
the independent measurement of T-odd and CPT-odd asymmetries,
but this route requires sources of electron neutrinos and antineutrinos above the muon mass energies, 
which is at present unavailable for accelerator facilities. 
As an alternative, our work consists in disentangling these two components, 
genuine and matter-induced CP violation, in the observable CP asymmetry.

Neutrino oscillations in matter are described through the effective
Hamiltonian in the flavor basis~\cite{MSW-W, Matter-Hamiltonian-Barger, Matter-Hamiltonian-Kuo, Matter-Hamiltonian-Zaglauer, Matter-Hamiltonian-Krastev, Matter-Hamiltonian-Parke}
\begin{align}
	\nonumber
	H &=
	\frac{1}{2E}
	\left\{
		U
	\mqty[
		m_1^2 &0 &0\\
		0 &m_2^2 &0\\
		0 &0 &m_3^2
	] 
	U^\dagger
	+
	\mqty[
		\,a\, &\,0\, &\,0\,\\
		0 &0 &0\\
		0 &0 &0
	] \right\}\\
	\label{eq:H}
	&=\frac{1}{2E}\; \tilde U \tilde M^2 \tilde U^\dagger\,,
\end{align}
where the first term describes neutrino oscillations in vacuum 
and the second one accounts for matter effects. 
The $a$ parameter is given by $a =2E V$, 
with $V$ the interaction potential and $E$ the relativistic neutrino energy.
For antineutrinos, $U \to U^*$ and $a \to -a$.
All neutrino masses $(\tilde M^2)$ and mixings $(\tilde U)$ in matter,
i.e. eigenvalues and eigenstates of $H$, 
can be calculated in terms of the parameters in the vacuum Hamiltonian 
$(M^2,\, U)$ and $a$.
Several analytical approaches have been presented in the 
literature~\cite{Analytical-Expressions-Cervera, Analytical-Expressions-Blennow, Analytical-Expressions-Parke, Analytical-Expressions-Ioannisian} using
approximations based on the hierarchical values of the different
parameters. For later purposes in this paper, we have developed~\cite{analytical-segarra}
a perturbative approach in $\dm_{21}$ --against $\dm_{31}$-- for
any value of 
$a$. 

The exact Hamiltonian leads to the flavor oscillation probabilities
\begin{align}
	\nonumber
	P(\nu_\alpha \to \nu_\beta)
	= &\, \delta_{\alpha\beta}
	-4\sum_{j<i}\mathrm{Re}~\tilde J_{\alpha\beta}^{ij}\,
	\sin^2 \tilde \Delta_{ij}\, -\\
	\label{eq:Pab}
	&-2\sum_{j<i}\mathrm{Im}~\tilde J_{\alpha\beta}^{ij}\,
	\sin 2\tilde \Delta_{ij}\,,
\end{align}
where 
$\tilde J_{\alpha\beta}^{ij} \equiv 
\tilde U_{\alpha i} \tilde U^*_{\alpha j}
\tilde U^*_{\beta i} \tilde U_{\beta j}$
are the rephasing-invariant mixings
and
$ \tilde \Delta_{ij} \equiv \frac{\Delta\tilde m^2_{ij} L}{4 E}$.
Notice that both 
$\tilde J_{\alpha\beta}^{ij}$
and
$\dmt_{ij} $
are energy dependent in matter.
Due to the CPV, CPTV $a \neq 0$ interaction,
neutrinos and antineutrinos acquire different masses and the complex
mixings do not satisfy the condition  
$\tilde {\bar J} = {\tilde J}^*$,
where
$\tilde {\bar J}$
refers to antineutrinos.
The T-odd genuine CP violation in matter needs, 
on the other hand,
$\tilde J \neq {\tilde J}^*$, $\tilde{\bar J} \neq \tilde{\bar J}^*$.
From these results on the different behavior under the discrete T and CPT symmetry transformations,
one derives the \textbf{Asymmetry Disentanglement Theorem} by separating the observable CP asymmetry 
in any flavor transition into $L$-even (CPT-violating) and $L$-odd (T-violating) functions,
\begin{align}
	\label{eq:Theorem}
	\mathcal{A}^\mathrm{CP}_{\alpha\beta} \equiv&\, 
	P(\nu_\alpha \to \nu_\beta) - 
	P(\bar \nu_\alpha \to \bar \nu_\beta) = 
	\mathcal{A}^\mathrm{CPT}_{\alpha\beta} +
	\mathcal{A}^\mathrm{T}_{\alpha\beta}\,,\\
	\nonumber
	\mathcal{A}^\mathrm{CPT}_{\alpha\beta} &=
		-4 \sum\limits_{j<i} \left[ 
			\mathrm{Re}~\tilde J_{\alpha\beta}^{ij}\, \sin^2 \tilde \Delta_{ij}
			-
			\mathrm{Re}~\tilde {\bar J}_{\alpha\beta}^{ij}\, \sin^2 \tilde {\bar \Delta}_{ij}
		\right]\,,\\
	\nonumber
	\mathcal{A}^\mathrm{T}_{\alpha\beta} &=
		-2\sum\limits_{j<i} \left[ 
			\mathrm{Im}~\tilde J_{\alpha\beta}^{ij}\, \sin 2\tilde \Delta_{ij}
			-
			\mathrm{Im}~\tilde {\bar J}_{\alpha\beta}^{ij}\, \sin 2\tilde {\bar \Delta}_{ij}
		\right]\,,
\end{align}
where 
$\mathcal{A}^\mathrm{CPT}_{\alpha\beta}$
is T-invariant  and vanishes when $a = 0$, 
whereas 
$\mathcal{A}^\mathrm{T}_{\alpha\beta}$ 
is CPT-invariant and vanishes when the flavor mixing is real,
corresponding to the CP phase of the PMNS matrix $\delta = 0,\, \pi$.
Eqs.~(\ref{eq:Pab}, \ref{eq:Theorem}) remain valid for extended
 models with more mass eigenstates and a rectangular mixing matrix.

For the three-neutrino model,
the unique rephasing invariant 
associated to genuine CP violation 
$\mathcal{\tilde J}$ 
is related to its value in vacuum
$\mathcal{J}=c_{12}c^2_{13}c_{23}s_{12}s_{13}s_{23}\sin\delta$
~\cite{Jarlskog}~by
\begin{equation}
	\label{eq:Scott}
	\mathcal{\tilde J} \equiv \mathrm{Im}~\tilde J_{e\mu}^{12}
	= \frac{\Delta m_{21}^2 \Delta m_{31}^2 \Delta m_{32}^2}%
	{\Delta \tilde m_{21}^2 \Delta \tilde m_{31}^2 \Delta \tilde m_{32}^2}\;
	\mathcal{J}\,,
\end{equation}

This connection~\cite{Harrison-Scott} between vacuum and matter is due to the fact that
the quantity characterizing CP violation in the flavor basis~\cite{CP-condition}
remains invariant for any flavor-diagonal interaction. The proportionality of
$\mathcal{\tilde J}$ with $\dm_{21}$ in Eq.(\ref{eq:Scott}) 
explains the absence of genuine CP violation 
in the vanishing limit of this parameter, even if there are three 
non-degenerate neutrino masses in matter. This transmutation from
masses in vacuum to mixings in matter was already observed in Ref.~\cite{matter0th},
where the limit $\dm_{21}=0$ in vacuum led to 
${\tilde U}_{e1}=0$. 
The complexity of $\tilde J^{ij}_{\alpha \beta}$ is entirely due to $\sin\delta$. 

\begin{figure}[b]
	\centering
	\includegraphics[width=\linewidth]{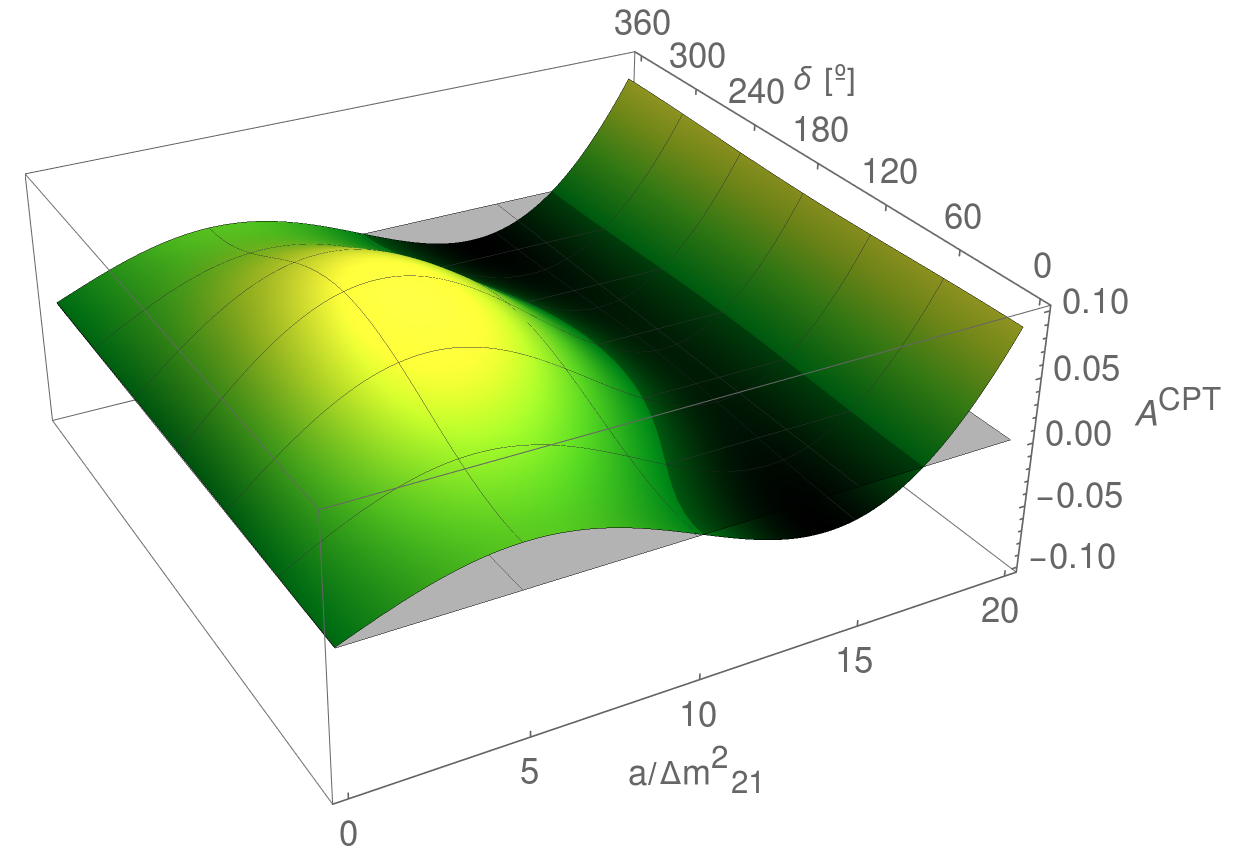}
	\hfill
	\includegraphics[width=\linewidth]{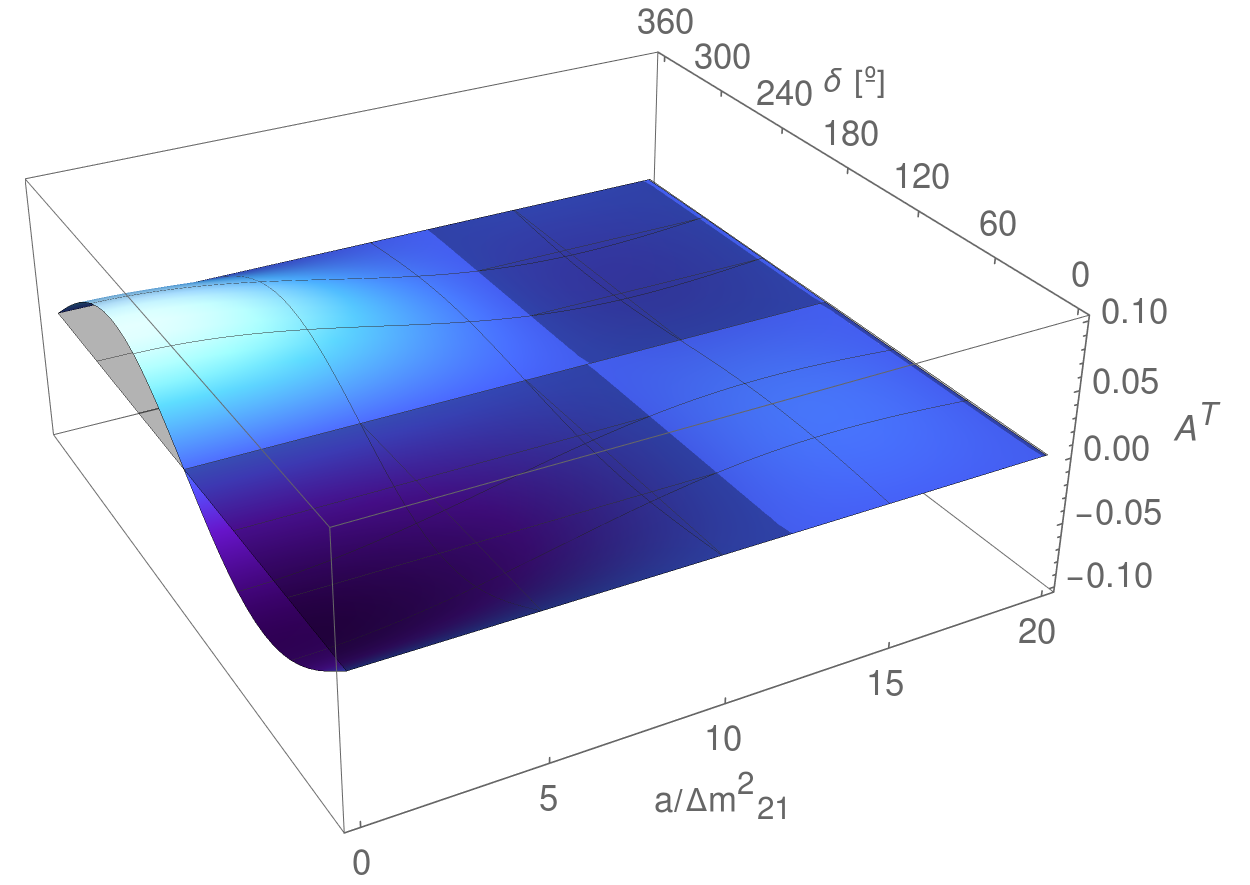}
	\caption{CPT-odd (up) and T-odd (down) components of 
	$\asym{CP}_{\mu e}$
	at $L = 1300$~km
	and $E = 0.75$~GeV as function of $a$ and $\delta$.
	}
	\label{fig:3Dasyms}
\end{figure}

We illustrate in Figure \ref{fig:3Dasyms} the power and expected behavior
of the Disentanglement Theorem by a separate representation of 
$\mathcal{A}^\mathrm{CPT}_{\mu e}$
and 
$\mathcal{A}^\mathrm{T}_{\mu e}$
as function of the matter potential $a$ and the CP phase $\delta$.
The chosen $(E,\, L)$ point gives comparable values of the two components, showing
their appropriate parities under these two parameters.

This and other results are produced
from a full numerical calculation
of the $\nu_\mu \to~\nu_e$ transition
produced by the exact Hamiltonian~(\ref{eq:H}),
using the best-fit mixing parameters 
from Ref.~\cite{mariam}
and assuming Normal Hierarchy,
unless otherwise specified.
Figure \ref{fig:3Dasyms} shows that 
$\mathcal{A}^\mathrm{CPT}_{\alpha\beta}$
is even in $\sin\delta\; \forall a$, 
and vanishes at $a = 0\; \forall \delta$, 
whereas 
$\mathcal{A}^\mathrm{T}_{\alpha\beta}$
is odd in $\sin\delta\; \forall a$, as requested by the theorem.

For neutrino propagation through the Earth mantle, valid for
terrestrial accelerator neutrinos, 
the value of the matter potential~\cite{a=3E}
allows us to write
$a~\approx~3 (E / \mathrm{GeV})\, \Delta m_{21}^2$.
In the energy region between the two MSW resonances,
$\Delta m_{21}^2 \ll a \ll \abs{\Delta m_{31}^2}$,
valid for both T2HK and DUNE neutrino energy spectra, 
one may approximate the relation in Eq.(\ref{eq:Scott}) between 
$\mathcal{\tilde J}$ in matter and $\mathcal{J}$ in vacuum as~\cite{analytical-segarra}
\begin{equation}
	\mathcal{\tilde J} 
	\approx
	\frac{\Delta m_{21}^2 \Delta m_{31}^2 }%
	{\abs{a} (1-\abs{U_{e3}}^2) \sqrt{(\Delta m_{31}^2-a)^2 + 
	4a \Delta m_{31}^2 \abs{U_{e3}}^2} }\,
	\mathcal{J}\,,
\end{equation}
where the proportionality factor is energy dependent through $a$.

\begin{figure}[b]
	\centering
	\includegraphics[width=\linewidth]{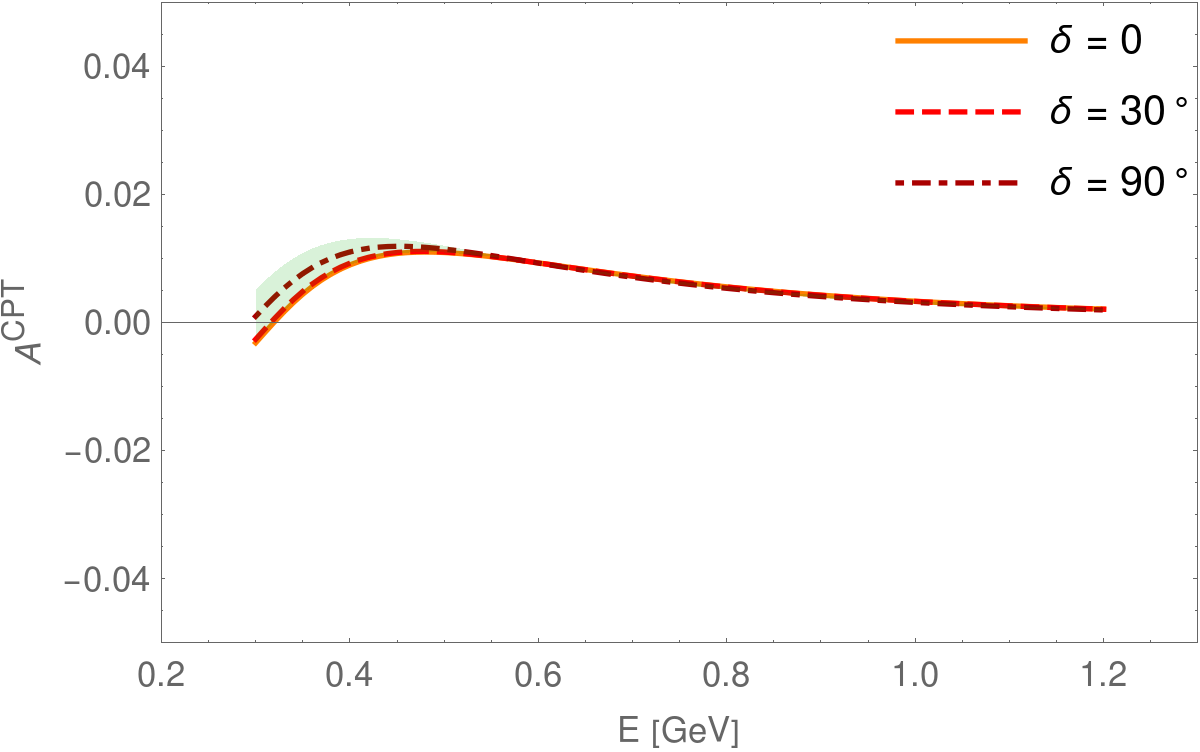}
	\hfill
	\includegraphics[width=\linewidth]{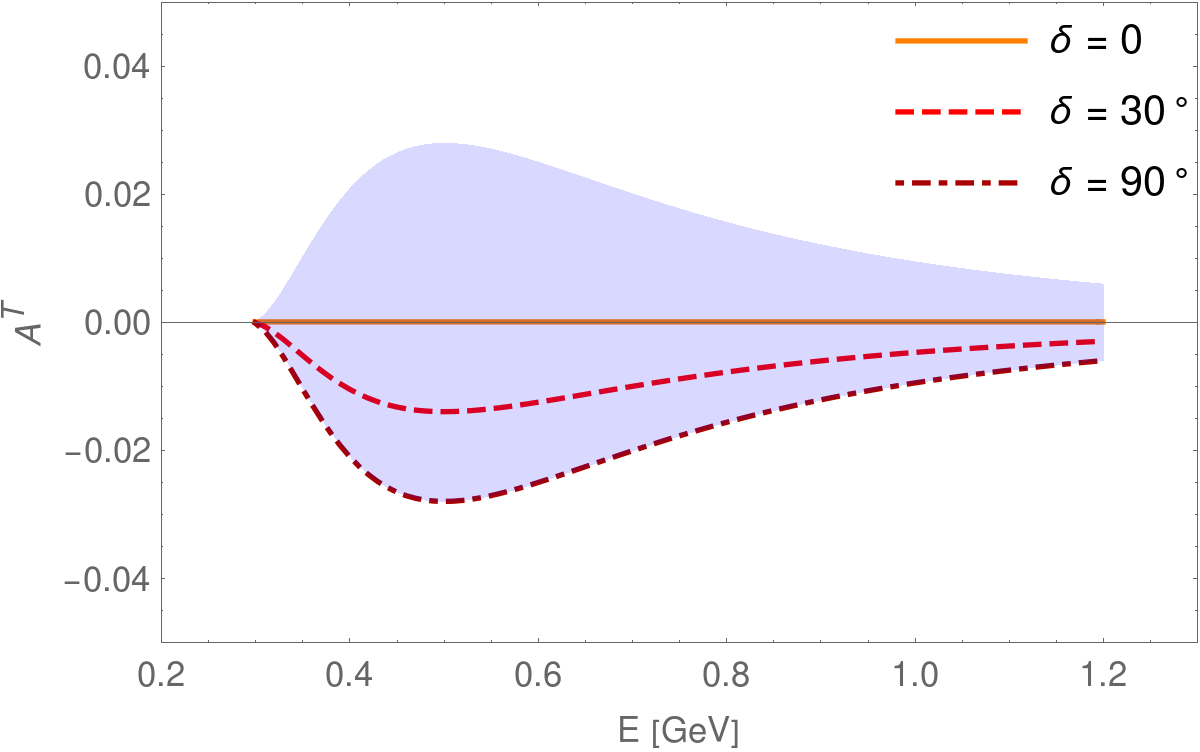}
	\caption{Energy distribution of the
		CPT-odd (up) and T-odd (down) components of the CP asymmetry 
	at T2HK.
	The bands correspond to all possible values changing $\delta$ in $(0,2\pi)$.
	}
	\label{fig:HK}
\end{figure}

In the following we explore whether, at fixed $L$ through the Earth mantle, 
the energy distributions of 
$\mathcal{A}^\mathrm{CPT}_{\mu e}$
and 
$\mathcal{A}^\mathrm{T}_{\mu e}$
present signatures of their separation in 
$\mathcal{A}^\mathrm{CP}_{\mu e}$
for T2HK and DUNE experiments.

Figure \ref{fig:HK} gives the predictions for the energy distributions of 
$\mathcal{A}^\mathrm{CPT}_{\mu e}$
and 
$\mathcal{A}^\mathrm{T}_{\mu e}$
at an intermediate baseline $L = 295$ km.
It is worth to note the lack of oscillating structure in the neutrino 
energy for both terms of the CP asymmetry. The magnitude of 
$\mathcal{A}^\mathrm{CPT}_{\mu e}$
is small, as expected, and slightly dependent on the CP phase $\delta$ through
the genuine CP-conserving small contributions of order $\dm_{21}$, more visible at low energies. 
When these results for a Normal Hierarchy are
re-calculated for an Inverted Hierarchy, the net effect is essentially
a change of sign in 
$\mathcal{A}^\mathrm{CPT}_{\mu e}$.
On the contrary, the magnitude of 
$\mathcal{A}^\mathrm{T}_{\mu e}$
is proportional to $\sin\delta$, as predicted by Eq.(\ref{eq:Scott}), without any
degeneracy when $\delta$ is varied in the entire interval from 0 to $2\pi$.
In addition, in this genuine term of the CP asymmetry the hierarchy
in the neutrino mass spectrum plays no role: its sign remains invariant 
when the sign of 
the largest mass splitting
is changed. We conclude that the sign of
$\mathcal{A}^\mathrm{CPT}_{\mu e}$
fixes the hierarchy whereas the magnitude and sign of 
$\mathcal{A}^\mathrm{T}_{\mu e}$
fixes $\sin\delta$. 

The beautiful different behavior of 
$\mathcal{A}^\mathrm{CPT}_{\alpha\beta}$
and
$\mathcal{A}^\mathrm{T}_{\alpha\beta}$
for the discrimination of the hierarchy is well understood to leading
order in $\dm_{21}$: 
zeroth order for 
$\mathcal{A}^\mathrm{CPT}_{\alpha\beta}$,
independent of $\delta$, 
and first order for 
$\mathcal{A}^\mathrm{T}_{\alpha\beta}$.
The mass spectrum in matter
changes from neutrinos to antineutrinos  as
\begin{equation}
	\Delta \tilde m_{21}^2 \leftrightarrow \Delta \tilde {\bar m}_{21}^2\,,
	\hspace{0.2cm}
	\Delta \tilde m_{31}^2 \leftrightarrow -\Delta \tilde {\bar m}_{32}^2\,,
	\hspace{0.2cm}
	\Delta \tilde m_{32}^2 \leftrightarrow -\Delta \tilde {\bar m}_{31}^2\,.
\end{equation}
Under this exchange of neutrinos by antineutrinos, the 
imaginary part of $\tilde J_{\alpha\beta}^{ij}$, as that of
$J_{\alpha\beta}^{ij}$, changes sign whereas the real parts do not. 
As the CP asymmetries are a difference between neutrino and antineutrino
oscillation probabilities, we discover that 
$\mathcal{A}^\mathrm{CPT}_{\alpha\beta}$
is changing its sign whereas the sign of
$\mathcal{A}^\mathrm{T}_{\alpha\beta}$
remains invariant under the change of hierarchy,
as seen in our numerical results.

\begin{figure}[t]
	\centering
	\includegraphics[width=\linewidth]{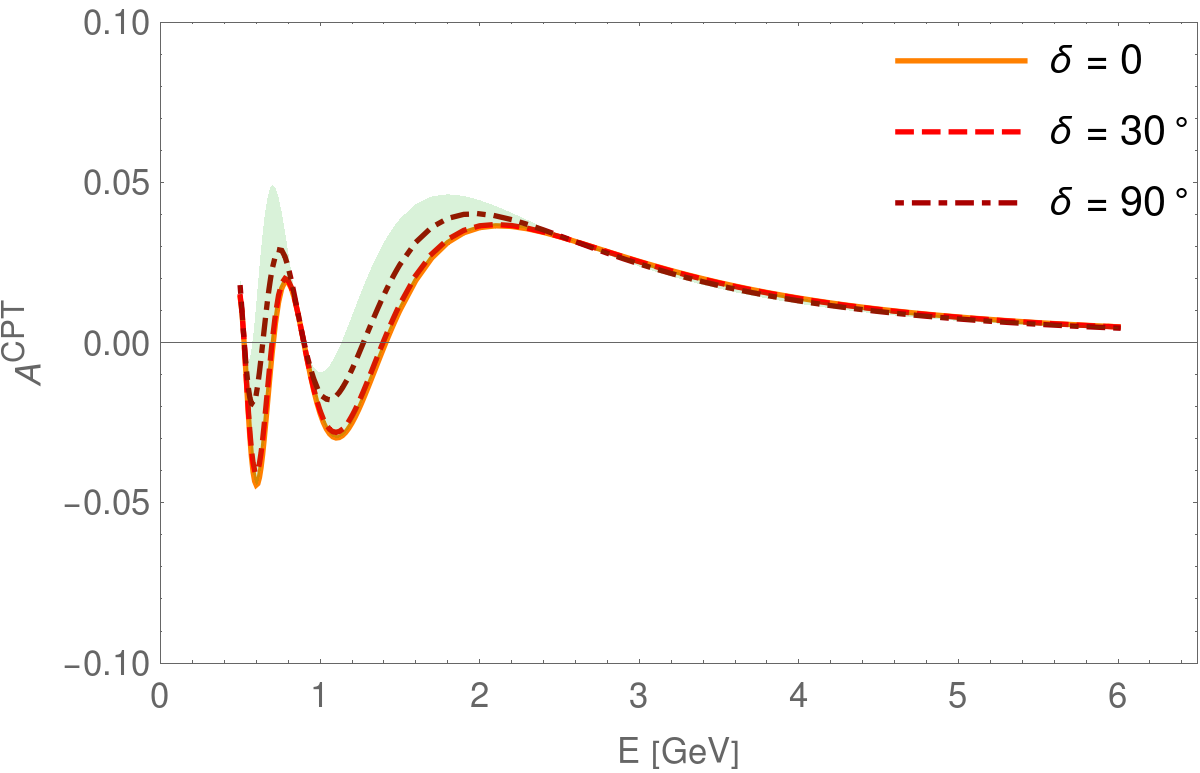}
	\hfill
	\includegraphics[width=\linewidth]{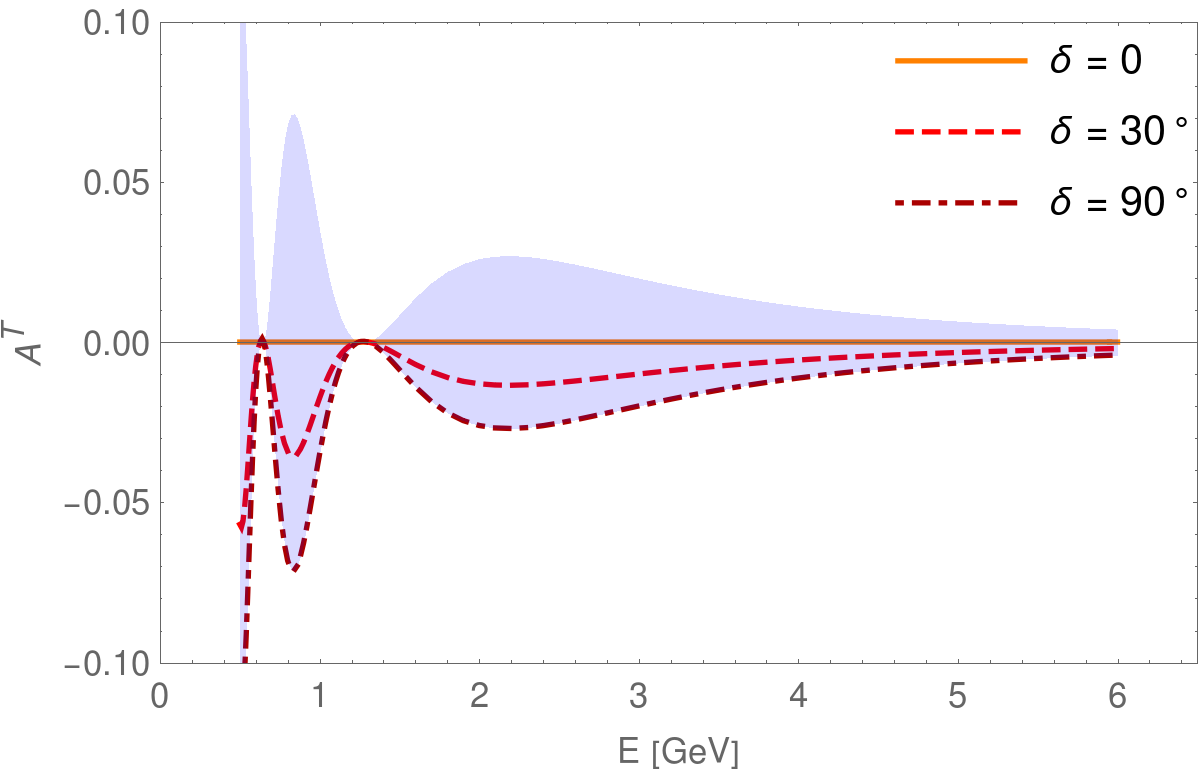}
	\caption{As Figure \ref{fig:HK} for the DUNE experiment.
		Notice a factor of 2 in the asymmetry scale with respect to Fig.\ref{fig:HK}.
	}
	\label{fig:Bands_DUNE}
\end{figure}

The increase in the baseline from $L = 295$ km to $L = 1300$ km has a very
important implication: the appearance of oscillations 
in the low and medium neutrino energy regions of the two distributions.
There is a different pattern for the two components
$\mathcal{A}^\mathrm{CPT}_{\mu e}$
and 
$\mathcal{A}^\mathrm{T}_{\mu e}$
of the experimental CP asymmetry,
with the zeros at different values and 
$\mathcal{A}^\mathrm{CPT}_{\mu e}$
changing its sign around the zeros, whereas
$\mathcal{A}^\mathrm{T}_{\mu e}$
does not.
This contrast is very well apparent in
the results we show in Fig.\ref{fig:Bands_DUNE}.

Besides this additional effect of having in the different energy distributions
a signature to separate the two components, all the other
properties discussed above remain the same, independent of the baseline.
These are the slight dependence of 
$\mathcal{A}^\mathrm{CPT}_{\mu e}$
on $\delta$
due to effects of $\dm_{21}$ at low energies and the hierarchy
discrimination with its sign,
as well as the proportionality of 
$\mathcal{A}^\mathrm{T}_{\mu e}$
with $\sin\delta$ independent of the neutrino hierarchy.

\begin{figure}[t]
	\centering
	\includegraphics[width=\linewidth]{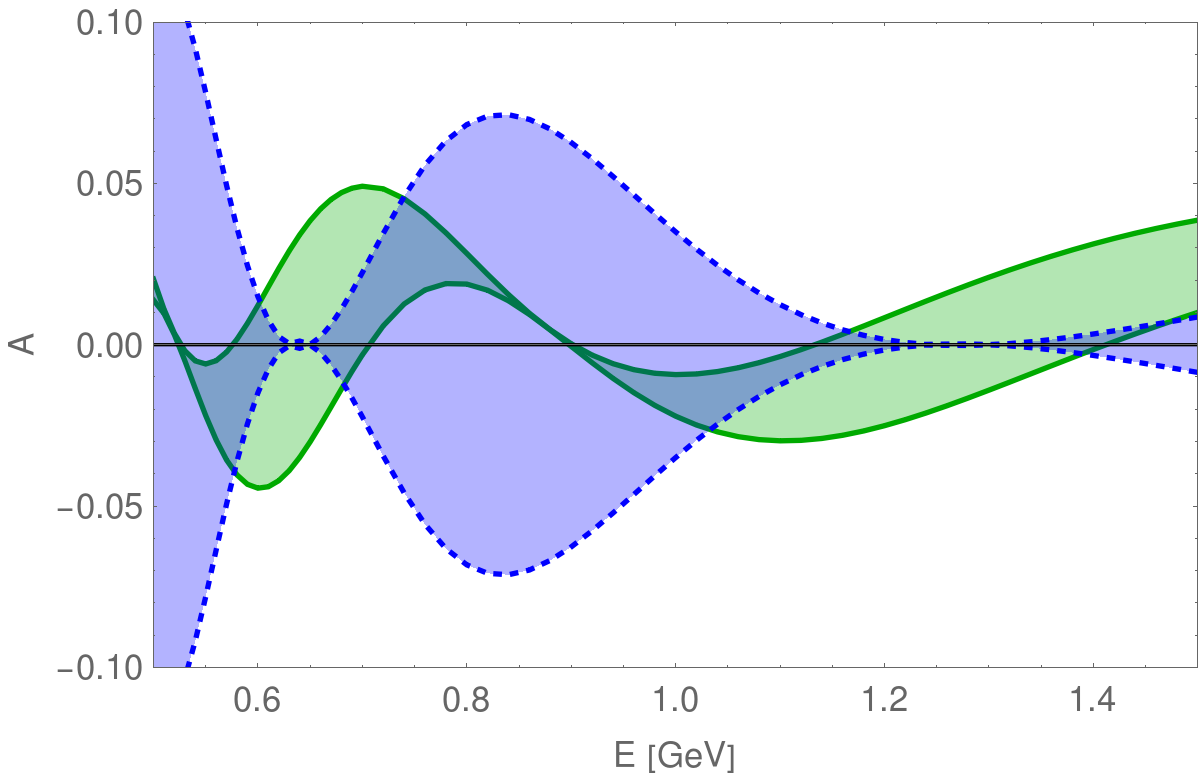}
	\caption{ Zooming Figure \ref{fig:Bands_DUNE} at low $E$,
		superposing the CPT-odd (green/solid) and T-odd (blue/dashed) 
		asymmetries.
	}
	\label{fig:zoom}
\end{figure}

In our scan of the different energy distribution behavior at $L = 1300$ km
of the matter-induced 
$\mathcal{A}^\mathrm{CPT}_{\mu e}$
and genuine 
$\mathcal{A}^\mathrm{T}_{\mu e}$
components of the CP asymmetry, 
we have discovered a magic energy interval around
$E = 0.91$ GeV with a zero for 
$\mathcal{A}^\mathrm{CPT}_{\mu e}$
and a relative maximum for 
$\left| \mathcal{A}^\mathrm{T}_{\mu e} \right|$.
We checked that this energy value
changes linearly with the vacuum $\abs{\dm_{31}}$
and is both Hierarchy-independent and blind to all other fit parameters.
This remarkable configuration is well seen in the results presented
in Fig.\ref{fig:zoom}, 
with the two bands for
$\mathcal{A}^\mathrm{CPT}_{\mu e}$
(green) and
$\mathcal{A}^\mathrm{T}_{\mu e}$
(blue).
The zero in 
$\mathcal{A}^\mathrm{CPT}_{\mu e}$
is independent of $\delta$, and
$\mathcal{A}^\mathrm{CPT}_{\mu e}$
changes its sign around this first-order zero,
whereas $\mathcal{A}^\mathrm{T}_{\mu e}$ 
has a maximal value proportional to $\sin\delta$.
This behavior ensures that
for a bin width up to the feasible~\cite{fdez-mtnez} $0.15$~GeV,
the mean value of $\mathcal{A}^\mathrm{CPT}_{\mu e}$
is below $10\%\, \abs{\mathcal{A}^\mathrm{T}_{\mu e}}_\mathrm{max}$,
whereas $\mathcal{A}^\mathrm{T}_{\mu e}$
is above $95\%$ of its peak value.

To summarize, in this work we have provided a conceptual basis 
for disentangling genuine from matter-induced CP violation in neutrino oscillations.
This is achieved 
by means of a theorem in which the experimental CP asymmetry 
contains two terms with different behavior under T and CPT symmetries. 
Whereas the matter-induced component is an even function of the baseline, 
the genuine CP violating term is odd in $L$, as for the case in vacuum. 
In this separation 
$\mathcal{A}^\mathrm{CP}_{\alpha\beta} =
\mathcal{A}^\mathrm{CPT}_{\alpha\beta} + 
\mathcal{A}^\mathrm{T}_{\alpha\beta}$
we have demonstrated that 
\begin{itemize}
	\item 
$\mathcal{A}^\mathrm{CPT}_{\alpha\beta}$
is even in $\sin\delta$ $\forall a$, and vanishes at $a = 0$ $\forall \delta$, 
whereas
$\mathcal{A}^\mathrm{T}_{\alpha\beta}$
is odd in $\sin\delta$ $\forall a$. 

\item
$\mathcal{A}^\mathrm{CPT}_{\alpha\beta}$,
which is slightly dependent on $\delta$ at low energies, 
changes its sign in going from the Normal to the Inverted Hierarchy whereas 
$\mathcal{A}^\mathrm{T}_{\alpha\beta}$
remains the same. 

\end{itemize}

For flavor-diagonal interactions with matter, 
the genuine CP violation condition in the flavor basis remains the same 
for the propagation in matter and in vacuum. 
This implies a well defined proportionality between 
the genuine CP-odd rephasing-invariant mixings for both cases, 
the proportionality factor given by energy-dependent neutrino masses in matter. 

The planned experiments for terrestrial neutrinos propagating through the Earth mantle 
look for the CP asymmetry in the golden channel 
$\nu_\mu \to \nu_e$
at fixed baselines $L$ and continuum energy spectra $E$ of the neutrino beam. 
In the search of appropriate signatures for the
separation of the two components of the CP asymmetry, 
we have studied 
the two energy distributions,
presenting peculiar differences at fixed long baseline:
\begin{itemize}
\item
The zeros are at different energies for 
$\mathcal{A}^\mathrm{CPT}_{\mu e}$
and
$\mathcal{A}^\mathrm{T}_{\mu e}$,
and their behavior around the zero is different: 
$\mathcal{A}^\mathrm{CPT}_{\mu e}$
changes sign whereas 
$\mathcal{A}^\mathrm{T}_{\mu e}$
does not. 

\item
In particular,
we have discovered a magic configuration around $E = 0.91$ GeV at $L = 1300$ km
which presents a zero for 
$\mathcal{A}^\mathrm{CPT}_{\mu e}$,
independent of $\delta$, 
and an extremal value for 
$\mathcal{A}^\mathrm{T}_{\mu e}$
proportional to $\sin\delta$.
This energy value depends linearly on $\abs{\dm_{31}}$.
Calculated for
$\abs{\dm_{31}} = (2.5 \pm 0.03)\times 10^{-3}~\mathrm{eV}^2$,
the magic energy is $E = (0.91 \pm 0.01)$~GeV.
At this Hierarchy-independent point, 
a measurement of $\mathcal{A}^\mathrm{CP}_{\mu e}$
with energy resolution up to $0.15$~GeV
directly probes genuine CP violation
in the lepton sector.

\item
For energies above 1.5 GeV, the sign of the CP asymmetry discriminates the neutrino hierarchy.
\end{itemize}

\begin{acknowledgments}
The authors would like to acknowledge fruitful discussions with 
Francisco Botella, Anselmo Cervera and Sergio Palomares.
This research has been supported by MINECO Project FPA 2017-84543-P, 
Generalitat Valenciana Project GV PROMETEO 2017-033 and 
Severo Ochoa Excellence Centre Project SEV 2014-0398. 
A.S. acknowledges the MECD support through the FPU14/04678 grant.
\end{acknowledgments}


\begin{thebibliography}{99}

\bibitem{atmospheric}
Y. Fukuda et al. [Super-Kamiokande Collaboration],
\href{https://doi.org/10.1103/PhysRevLett.81.1562}{Phys. Rev. Lett. {\bf 81}, 1562 (1998)}.
\arxiv{hep-ex/9807003}

\bibitem{solar}
Q.R. Ahmad et al. [SNO Collaboration],
\href{https://doi.org/10.1103/PhysRevLett.89.011301}{Phys. Rev. Lett. {\bf 89}, 011301 (2002)}.
\arxiv{nucl-ex/0204008}
	
\bibitem{reactor}
T. Araki et al. [KamLAND Collaboration],
\href{https://doi.org/10.1103/PhysRevLett.94.081801}{Phys. Rev. Lett. {\bf 94}, 081801 (2005)}.
\arxiv{hep-ex/0406035}
	
\bibitem{accelerator}
Ahn et al. [K2K Collaboration],
\href{https://doi.org/10.1103/PhysRevD.74.072003}{Phys. Rev. {\bf D74}, 072003 (2006)}.
\arxiv{hep-ex/0606032}

\bibitem{mariam}
P.F. de Salas, D.V. Forero, C.A. Ternes, M. Tortola and F.W.F. Valle,
\href{https://doi.org/10.1016/j.physletb.2018.06.019}{Phys.Lett. {\bf B782}, 633 (2018)}.
\arxiv{1708.01186v2} [hep-ph]

\bibitem{nufit}
I. Esteban, M. C. Gonzalez-Garcia, M. Maltoni, I. Martinez-Soler and T. Schwetz,
\href{https://doi.org/10.1007/JHEP01(2017)087}{JHEP {\bf 01}, 087 (2017)}.
\arxiv{1611.01514} [hep-ph]

\bibitem{fogli}
F. Capozzi, G.L. Fogli, E. Lisi, A. Marrone, D. Montanino and A. Palazzo,
Phys.Rev. {\bf D89}, 093018 (2014).
\href{https://doi.org/10.1103/PhysRevD.89.093018}{Phys.Rev. {\bf D89}, 093018 (2014)}.
\arxiv{1312.2878} [hep-ph]

\bibitem{HK}
K. Abe et al. [Hyper-Kamiokande Proto-Collaboration],
\href{http://www.hyperk.org/?p=215}{KEK-Preprint-2016-21, ICRR-Report-701-2016-1}.


\bibitem{DUNE}
R. Acciarri et al. [DUNE Collaboration],
FERMILAB-DESIGN-2016-02.
\arxiv{1512.06148} [physics.ins-det]

\bibitem{T2K}
K. Abe et al. [T2K Collaboration], 
\href{https://doi.org/10.1103/PhysRevLett.107.041801}{Phys.Rev.Lett. {\bf 107}, 041801 (2011)}.
\arxiv{1106.2822} [hep-ex]

\bibitem{2CHOOZ}
Y. Abe et al. [Double Chooz Collaboration], 
\href{https://doi.org/10.1103/PhysRevLett.108.131801}{Phys.Rev.Lett. {\bf 108}, 131801 (2012)}.
\arxiv{1112.6353} [hep-ex]

\bibitem{DAYABAY}
F. An et al. [Daya Bay Collaboration], 
\href{https://doi.org/10.1103/PhysRevLett.108.171803}{Phys.Rev.Lett. {\bf 108}, 171803 (2012)}.
\arxiv{1203.1669} [hep-ex]

\bibitem{RENO}
J.K. Ahn et al. [RENO Collaboration], 
\href{https://doi.org/10.1103/PhysRevLett.108.191802}{Phys.Rev.Lett. {\bf 108}, 191802 (2012)}.
\arxiv{1204.0626} [hep-ex]

\bibitem{baryogenesis}
M. Fukugita and T. Yanagida,
\href{https://doi.org/10.1016/0370-2693(86)91126-3}{Phys.Lett. {\bf B174}, 45 (1986)}.

\bibitem{MSW-W}
L. Wolfenstein,
\href{https://doi.org/10.1103/PhysRevD.17.2369}{Phys.Rev.D {\bf 17}, 2369 (1978)}.

\bibitem{MSW-MS}
S.P. Mikheyev and A.Yu. Smirnov,
Sov.J.Nucl.Phys. {\bf 42}, 913 (1985).

\bibitem{obs:donini}
A. Donini, M.B. Gavela, P. Hernandez and S. Rigolin,
\href{https://doi.org/10.1016/S0550-3213(00)00029-8}{Nucl.Phys. {\bf B574}, 23 (2000)}.
\arxiv{hep-ph/9909254}

\bibitem{obs:akhmedov}
E.Kh. Akhmedov, M. Maltoni and A.Yu. Smirnov,
\href{https://doi.org/10.1088/1126-6708/2008/06/072}{JHEP {\bf 0806}, 072 (2008)}.
\arxiv{0804.1466} [hep-ph]

\bibitem{obs:ohlsson}
T. Ohlsson, H. Zhang and S. Zhou,
\href{https://doi.org/10.1103/PhysRevD.87.053006}{Phys.Rev. {\bf D87}, 053006 (2013)}.
\arxiv{1301.4333} [hep-ph]

\bibitem{Banuls}
J. Bernabeu and M.C. Banuls,
\href{https://doi.org/10.1016/S0920-5632(00)00690-3}{Nucl.Phys.Proc.Suppl. {\bf 87}, 315 (2000)}.
\arxiv{hep-ph/0003299}
	

\bibitem{Matter-Hamiltonian-Barger}
V. Barger, K. Whisnant, S. Pakvasa and R.J.N. Phillips, 
\href{https://doi.org/10.1103/PhysRevD.22.2718}{Phys.Rev.D {\bf 22}, 2718 (1980)}.

\bibitem{Matter-Hamiltonian-Kuo}
T.K. Kuo and J. Pantaleone, 
\href{https://doi.org/10.1103/RevModPhys.61.937}{Rev.Mod.Phys. {\bf 61}, 937 (1989)}.

\bibitem{Matter-Hamiltonian-Zaglauer}
H.W. Zaglauer and K.H. Schwarzer, 
\href{https://doi.org/10.1007/BF01555889}{Z.Phys. {\bf C40}, 273 (1988)}.

\bibitem{Matter-Hamiltonian-Krastev}
P. Krastev, 
\href{https://doi.org/10.1007/BF02790019}{Nuovo Cim. {\bf A103}, 361 (1990)}.

\bibitem{Matter-Hamiltonian-Parke}
R.H. Bernstein and S.J. Parke, 
\href{https://doi.org/10.1103/PhysRevD.44.2069}{Phys.Rev.D {\bf 44}, 2069 (1991)}.

\bibitem{Analytical-Expressions-Cervera}
A. Cervera, A. Donini, M.B. Gavela, J.J. Gomez Cadenas, P. Hernandez, O. Mena and S. Rigolin,
\href{https://doi.org/10.1016/S0550-3213(00)00221-2}{Nucl.Phys. {\bf B593}, 731 (2001)}.
\arxiv{hep-ph/0002108}

\bibitem{Analytical-Expressions-Blennow}
M. Blennow and A.Yu. Smirnov, 
\href{https://doi.org/10.1155/2013/972485}{Adv.High Energy Phys. {\bf 2013}, 972485 (2013)}.
\arxiv{1306.2903} [hep-ph]

\bibitem{Analytical-Expressions-Parke}
P.B. Denton, H. Minakata and S.J. Parke,
\href{https://doi.org/10.1007/JHEP06(2016)051}{JHEP {\bf 1606}, 051 (2016)}.
\arxiv{1604.08167} [hep-ph]

\bibitem{Analytical-Expressions-Ioannisian}
A. Ioannisian and S. Pokorski,
\arxiv{1801.10488} [hep-ph]

\bibitem{analytical-segarra}
J. Bernabeu and A. Segarra, 
\arxiv{1807.11879} [hep-ph]


\bibitem{Jarlskog}
C. Jarlskog,
\href{https://doi.org/10.1007/BF01565198}{Z.Phys. {\bf C29}, 491 (1985)}.

\bibitem{Harrison-Scott}
P.F. Harrison and W.G. Scott,
\href{https://doi.org/10.1016/S0370-2693(00)00153-2}{Phys.Lett. {\bf B476}, 349 (2000)}.
\arxiv{hep-ph/9912435}

\bibitem{CP-condition}
J. Bernabeu, G.C. Branco and M. Gronau,
\href{https://doi.org/10.1016/0370-2693(86)90659-3}{Phys.Lett. {\bf B169}, 243 (1986)}.

\bibitem{matter0th}
M.C. Banuls, G. Barenboim and J. Bernabeu,
\href{https://doi.org/10.1016/S0370-2693(01)00723-7}{Phys.Lett. {\bf B513}, 391 (2001)}.
\arxiv{hep-ph/0102184}

\bibitem{a=3E}
I. Mocioiu and R. Shrock, 
\href{https://doi.org/10.1103/PhysRevD.62.053017}{Phys.Rev. {\bf D62}, 053017 (2000)}.
\arxiv{hep-ph/0002149}

\bibitem{fdez-mtnez}

V. De Romeri, E. Fernandez-Martinez and M. Sorel,
\href{https://doi.org/10.1007/JHEP09(2016)030}{JHEP {\bf 1609}, 030 (2016)}.
\arxiv{1607.00293} [hep-ph]

\end{thebibliography}
\end{document}